# A PIONIER View on Mass-Transferring Red Giants

Henri M.J. Boffin[1], Nicolas Blind[2], Michel Hillen[3], Jean-Philippe Berger[1], Alain Jorissen[4], Jean-Baptiste Le Bouquin[5]

1 ESO
2 Max Planck Institut für extraterrestrische Physik, Garching, Germany
3 Instituut voor Sterrenkunde, Katholiek Universiteit Leuven, Belgium
4 Institut d'Astronomie et d'Astrophysique, Université Libre de Bruxelles, Belgium
5 UJF-Grenoble CNRS-INSU, Institut de Planétologie et d'Astrophysique de Grenoble, France

**Symbiotic stars display absorption lines of a cool red giant together with emission lines of a nebula ionized by a hotter star, indicative of an active binary star system in which mass transfer is occurring. PIONIER at the VLT has been used to combine the light of four telescopes at a time to study in unprecedented detail how mass is transferred in symbiotic stars. The results of a mini-survey of symbiotic stars with PIONIER are summarised and some tentative general results about the role of Roche lobe overflow are presented.**

Symbiotic stars are a class of bright, variable red giant stars, which show a composite spectrum: on top of the typical absorption features of the cool star, there are strong hydrogen and helium emission lines, linked to the presence of a hot star and a nebula. It is now well established that such a "symbiosis" is linked to the fact that these stars are active binary systems, with orbital periods between a hundred days and several years. On account of the variable nature of these systems and their clear signs of accretion (many are often associated with "novae" or undergo outbursts of some sort), it is usually accepted that the red giant is losing mass that is partly transferred to the accreting companion – either a main-sequence (MS) star or a white dwarf (WD); as developed by Kenyon & Webbink (1984) and Mürset et al. (1991). Symbiotic stars are *de facto* the low- and intermediate-mass interacting binaries with the longest orbital periods and the largest component separations. They are thus excellent laboratories to study a large spectrum of very poorly understood physical processes, with wide ranging applications (Mikolajewska, 2007). Symbiotic stars containing a white dwarf may also be progenitors of Type Ia supernovae.

## The mass transfer process

One of the main questions related to symbiotic stars is how the mass transfer takes place: by stellar wind, through Roche-lobe overflow (RLOF) or through some intermediate process? Red giants naturally loose mass through a stellar wind, with typical mass-loss rates of $10^{-8}$ $M_{sun}$/yr on the first ascent of the red giant branch, and up to $10^{-6}$ $M_{sun}$/yr, or more, on the asymptotic giant branch (AGB). The wind velocity is rather low, about 5–30 km/s in most cases, and part of this wind can thus be accreted by the companion. Moreover, the red giants in symbiotic systems are known to have stronger mass-loss rates than normal red giants. Although still not understood, this means that mass transfer can be even stronger.

Alternatively, if the system is a close enough binary, the red giant may fill its Roche lobe and start mass transfer through RLOF. However RLOF by a red giant with its huge convective envelope is quite problematic. Unless the mass ratio (initially greater than one, as it is the more massive star that evolves first into a giant) is smaller than some critical value (between about 1 and 1.5), the mass transfer will be dynamically unstable, leading to a common envelope (CE) phase, whose inescapable outcome is a very short (a few days or less) binary system. Such a process is able to explain short-period binaries such as cataclysmic variables or some central stars of planetary nebulae, but would not explain the long, observed periods of symbiotic stars. Thus, it is generally assumed that for RLOF to take place, and be stable so as to avoid the CE, the giant primary star will need to lose some mass by a stellar wind first (in order to bring the mass ratio down).

We now know about 200 symbiotic stars, but have orbital elements for only 40 systems or so. There are, however, also similar systems to symbiotic stars, i.e., binaries with a red giant primary but with lower mass-loss or mass transfer. Until very recently, it was not possible to firmly establish whether the mass-loss process in symbiotic and related stars took place via a wind or through RLOF. Answering this question indeed requires being able to compare the radius of the stars to the Roche lobe radius (which depends on the star separation and the mass ratio). However, determining the radius of the red giant in symbiotic systems is not straightforward, and there has always been some controversy surrounding these determinations.

During the last decade, clear evidence of ellipsoidal variations in the near-infrared light curves of many symbiotic systems has been presented, suggesting that the giants are filling their Roche lobes, or are at least very close to doing so. The radii estimated from the light curves, however, lead to a continually embarrassing problem: they are systematically discrepant with radii derived from rotation velocities by a factor of two (Mikolajewska, 2007)! Several explanations have been proposed to account for this discrepancy, but the only way to answer it is by measuring the giant's radius for symbiotic stars in a direct way!

Optical interferometry is currently the only available technique that can achieve this aim. Interferometry allows the size and distortion of the giant star, and in some cases, the orbital parameters of the system to be determined, without any a priori estimate on their characteristics. This unique ability has already allowed us to study in unprecedented detail the interacting binary system SS Leporis, leading to a complete revision of the system (Blind et al., 2011).

## Revisiting SS Lep

SS Leporis is composed of an evolved M giant and an A star in a 260-day orbit, and presents the most striking effect of mass transfer, the "Algol" paradox; that is, the M giant is less massive than the A star. The A star is unusually luminous and surrounded by an expanding shell, certainly as the result of accretion. The observation of regular outbursts and of ultraviolet activity from the A star shell are further hints that mass transfer is currently ongoing. Additionally, a large circumbinary dust disc surrounds the binary system, implying that the mass transfer process is non-conservative.

We observed SS Lep during the commissioning of the PIONIER visitor instrument at the Very Large Telescope Interferometer (Le Bouquin et al. 2011). Since it combines four telescopes (in our case, the four 1.8-metre Auxiliary Telescopes), PIONIER provides six visibilities and four closure phases simultaneously, which allows reliable image reconstruction. We were able to detect the two components of SS Lep as they moved across their orbit (see Figure 1) and to measure the diameter of the red giant in the system (~2.2 milli-arcseconds, mas). By reconstructing the visual orbit and combining it with the previous spectroscopic one, it was possible to well constrain the parameters of the two stars.

The M giant is found to have a mass of 1.3 $M_{sun}$, while the less evolved A star has a mass twice as large: thus a clear mass reversal must have taken place, with more than 0.7 $M_{sun}$ having been transferred from one star to the other. Our results also indicate that the M giant only fills around 85 +/- 3% of its Roche lobe, which means that the mass transfer in this system is most likely by a wind and not by RLOF. However, as the M giant is currently thought to be in the early-AGB phase, where mass loss is still very small, it is difficult to understand how it could have lost so much mass in a few million years (the time spent on the AGB), unless one invokes the companion-reinforced attrition process (CRAP) of Tout & Eggleton (1988), i.e. assuming that the tidal force of the companion dramatically increases the wind mass loss. Although this process allows, in principle, to the current state of SS Lep to be explained without resorting to a Roche-lobe overflow, the validity of this assumption still needs to be proved by a more detailed study.

## The PIONIER mini-survey

It is not always possible to obtain as many data points as in the case of SS Lep (e.g., in most symbiotic systems, the companion would not have a detectable signature in the infrared) and thus to constrain the system to the same extent. However, there are several systems for which it is possible to determine, with great accuracy, the diameter of the mass-losing giant. Thus combined with previous data, the Roche lobe-filling factor could already be constrained. Recently we observed with PIONIER several symbiotic and related stars, measuring their diameters – in the range 0.6 to 2.3mas – and thereby assessing for the first time, in an independent way, the filling factor of the Roche lobe of the mass-losing giants (Boffin et al., 2014).

For the three stars with the shortest orbital periods (i.e., HD 352, FG Ser and HD 190658), we find that the giants are filling (or are close to filling) their Roche lobes, consistent with the fact that these objects present ellipsoidal variations in their light curves. In the case of the symbiotic star FG Ser, we find that the diameter is changing by 13% over the course of 41 days, while observations of HD352 are indicative of an elongation at the level of 10% (see Figure 2). Such deformations need, however, to be confirmed with more precise interferometric observations and compared with the outcome of light-curve modelling. The other three studied stars (V1261 Ori, ER Del, and AG Peg) have filling factors in the range 0.2 to 0.55, i.e., the star is well within its Roche lobe.

## A possible dichotomy?

We here tentatively propose that the three systems which apparently fill their Roche lobes (HD 352, FG Ser and HD 190658) are those that contain a main-sequence companion or a helium (He) white dwarf (WD), and not a carbon/oxygen (CO) WD. It is for example very useful to compare the systems FG Ser and V1261 Ori. Despite the similarity of their orbital period (P=633.5$^d$ and 638.2$^d$, respectively), and the not so different physical properties (viz., temperature and luminosity of the red giant), the former has most likely a companion with too low a mass to allow for a CO WD, while the latter has a CO WD companion, given its pollution in s-process elements (Ake et al., 1991). We found that FG Ser is a nominal case of a synchronous giant filling its Roche lobe, while the red giant in V1261 Ori is filling only about 30% of its Roche lobe and mass transfer occurs by a wind. The fact that in FG Ser the giant has a rotation synchronised with the orbital motion and a circular orbit (both properties expected for a Roche-lobe filling system), while V1261 Ori is clearly not synchronised and has still a slightly eccentric orbit (e=0.07), is another illustration of this possible dichotomy (see Table 1).

This conclusion on the Roche lobe filling would provide some credit to the work of Kenyon & Webbink (1984) who concluded that, in order to explain the properties of symbiotic stars, one needed RLOF (or quasi-RLOF) in systems with main-sequence companions, but wind mass loss in systems with white dwarfs. SS Lep, which almost fills its Roche lobe, and which has a clear main sequence companion, also fits the picture. It is, however, still difficult to understand why, if Roche lobe overflow takes place in our three systems, it does not lead to a dynamic CE, as the mass ratio we determine for our Roche-lobe filling giants are often larger than 1.5. Clearly more theoretical work is needed along those lines.

Our detailed analysis of these stars also allows us to place the systems in an HR-diagram (Figure 3). The analysis has shown that red giants in symbiotic systems are rather normal and obey similar relations between colour, spectral type, temperature, luminosity and radius. Thus, the fact that they have larger mass-loss rates than single giants must be linked in some way to their binarity, adding credence to the CRAP mechanism.

The observations presented here clearly reveal the power of interferometry for the study of interacting binary stars. The major limitation to our study is the rather imprecise knowledge of the distance of these objects. The GAIA survey, soon to begin, will be most useful in this respect.

**Table 1:** A possible dichotomy among symbiotic stars?

| Companion star | CO WD | He WD or MS |
|---|---|---|
| Red giant shows | s-process enhancement | no s-process enhancement |
| Rotation of giant | non-synchronised | synchronised |
| Eccentricity | e > 0 | e=0 |
| Mass transfer | Wind accretion | RLOF |

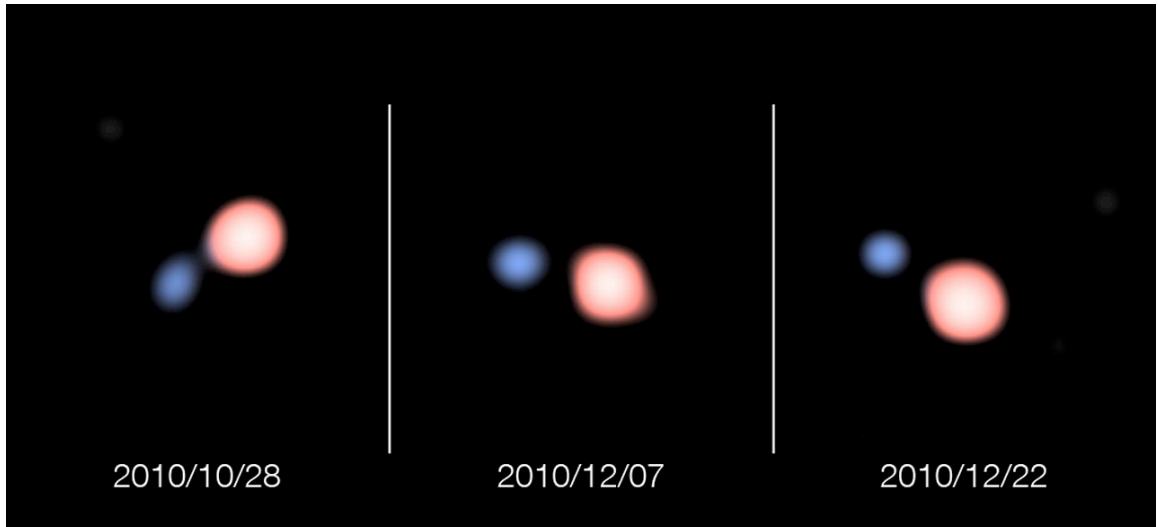

**Figure 1:** Image reconstruction of the PIONIER observations of SS Lep showing the orbital motion of the resolved M giant and the A star. The images are centered on the centre of mass. The distortion of the giant in the images is most certainly an artifact of the asymmetric PSF. See Release eso1148 for more details. Credit: ESO/PIONIER/IPAG.

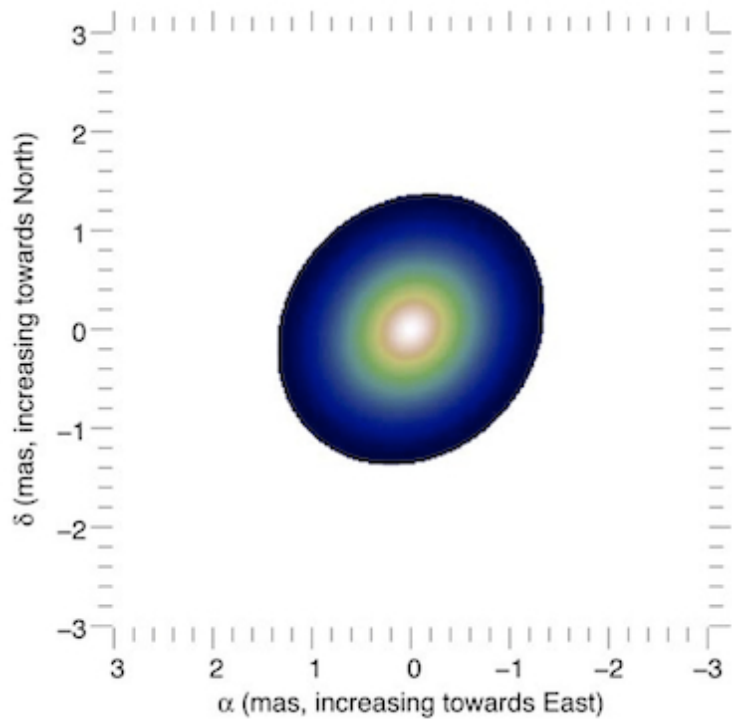

**Figure 2:** Model image of the elongated Gaussian that best fits the PIONIER visibilities for HD 352: the elongation ratio is 1.156 ± 0.026, the star is 1.38 × 1.6 mas wide, with the major axis position angle at 138 ± 4 degrees. Reproduced from Boffin et al. (2014).

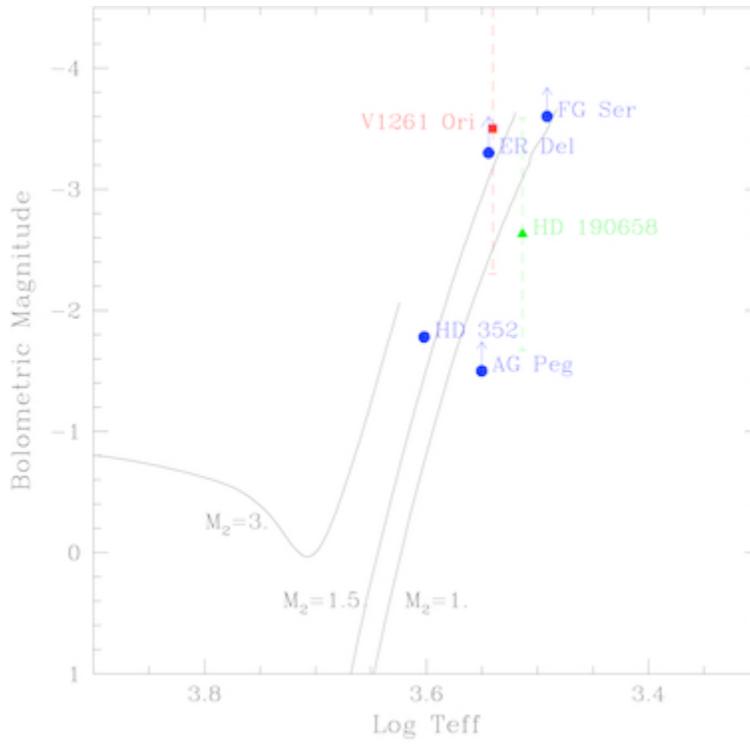

**Figure 3:** Hertzsprung-Russell diagram showing the positions of our target stars (full dots, in blue), together with Yonsei-Yale evolutionary tracks for solar abundance stars with initial masses of 1, 1.5, and 3 $M_{sun}$. For FG Ser, ER Del, and AG Peg, we have only lower limits for the bolometric magnitude. The range for V1261 Ori is indicated with the red dashed line. Adapted from Boffin et al. (2014).